\documentclass[12pt,a4paper]{article}
\usepackage{epsfig,amsfonts}
\newcommand{\be}{\begin{equation}}
\newcommand{\ee}{\end{equation}}
 
\newcommand{\qcsq}{\overline{\langle q_c^2 \rangle}} 
\newcommand{\qcfo}{\overline{\langle q_c^4 \rangle}} 

\begin{document}

\title{Study of a microcanonical algorithm on the \\
$\pm J$ spin glass model in $d=3$.}

\author{
J.J. Ruiz-Lorenzo$^a$ and C.L. Ullod$^b$\\[0.5em]
$^a$ {\small Departamento de F\'{\i}sica Te\'orica I, Universidad Complutense de Madrid,}\\
{\small Ciudad Universitaria, Madrid (Spain)}\\[0.3em]
$^b$ {\small Departamento de F\'{\i}sica Te\'orica, Universidad de Zaragoza,}\\
{\small Plaza S. Francisco s/n, Zaragoza (Spain)}\\[0.3em]
{\small \tt ruiz@lattice.fis.ucm.es}\\
{\small \tt clu@sol.unizar.es}\\[0.5em]}

\date{July 5, 1999}

\maketitle


\begin{abstract}

We consider a microcanonical local algorithm to be applied on the $\pm
J$ spin glass model. We have compared the results coming from a
microcanonical Monte Carlo simulation with those from a canonical one:
Thermalization times, spin glass susceptibilities and Binder
parameters. For a fixed lattice size we found different results
between the two thermodynamic ensembles, which tend to vanish at
bigger volumes. Moreover, microcanonical thermalization times are
longer than the canonical ones. Finally we have checked that one of the
Guerra relations is satisfied with good precision for the two largest
lattices.

\end{abstract}

\thispagestyle{empty}

\newpage


\section{Introduction}

The Ising spin glass is the paradigm of the complex systems. It possesses
two fundamental characteristics: disorder, because the couplings are
random variables and frustration since the signs of the coupling are
positive or negative.  These two characteristics produce the main
property of a complex system: a very slow dynamics.

This slow dynamics is due to the existence and competition of a large
number of pure and metastable states below the critical
temperature. In some cases, a large number of metastable states above
the critical temperature can produce this effect even in the paramagnetic 
phase.   

During the last two decades, the existence of a low-temperature phase
in Ising spin glass in three dimensions has been investigated and it
has consumed a large amount of CPU resources. At equilibrium it is
possible to simulate up to $L=16$ lattices and the signature of the
transition is very weak~\cite{YOUNG,MAPARU,INMAPARU}.

In reference~\cite{MAPARURI} was found a behaviour for the dynamical
critical exponent $z(T) \simeq 7 T_c/T$, where $T_c \simeq 1$ means
for the critical temperature and $T$ is the temperature, for the three
dimensional Gaussian Ising spin glass using a Metropolis
dynamics. This numerical fact has been corroborated in experiments
using samples of CuMn (at 6\%) and thiospinel~\cite{JOWHV}.  This
result for the dynamical critical exponent implies  Monte Carlo (MC)
thermalization times proportional to $L^7$ near the phase transition ($L$
is the size of the system). One can compare this behaviour with that of
the pure Ising model: the thermalization time diverges as $L^2$ (near
its critical temperature).

From the previous discussion follows that traditional approaches using
only  local algorithms should fail in thermalizing a large system.
Moreover, the absence of a non local update algorithm and the high
value of the dynamical critical exponent for the local ones convert
this problem in a very challenging computational issue. Recent works
using large amounts of computational power with a standard local MC
simulation \cite{YOUNG} showed some evidences of a cold phase in
this model.  Other recent approaches \cite{BOOK,ENZO} in similar
models have used more sophisticated update algorithms, based on a
combination of a standard local MC run and an innovative update
process in the temperature. These algorithms~\cite{BOOK,ENZO}, the
simulated tempering and parallel tempering, succeed in thermalizing
systems at temperatures lower than those reachable by a standard
Metropolis simulation.

Moreover, in a canonical simulation, the most time-consuming task 
is the generation of the random numbers, and so,
one possibility is to use a MC method that does not use random
numbers. This calls for microcanonical methods. In particular, 
Creutz developed the so-called demon algorithm that does not 
need random number generation to work.

The aim of this paper is to investigate the behavior of this
microcanonical local update algorithm on this model. Since that, we
will use only this algorithm, although it is clear that other tools,
as parallel tempering, have to be implemented to improve the
simulation in order to try to elucidate the low temperature regime of
this model.  In particular, we will study numerically the ergodicity
of the algorithm, efficiency (i.e. autocorrelation times) and the
difference with the canonical algorithm when we work at finite volume
and how these differences go to zero with the volume. It is clear,
that a study of this kind is essential if we will use the
microcanonical algorithm in extensive numerical simulations. 

As further studies we will plan to analyze the performance of a
combination of microcanonical and canonical algorithms. This
combination has worked very well in the simulation of some physical
systems as Quantum Chromodynamics~\cite{overrelax} and it could be of
great interest to check if this combination will work well in spin
glasses.

Moreover, one of the authors of this paper is finalizing the
construction of a dedicated machine to simulate this model \cite{CLU},
being this paper a preliminary study of the characteristics of the
demon algorithm, previous to its hardware implementation.  We remark
that this algorithm could also be used on computers of general purpose
not only in dedicated machines.


\section{Model, observables and update algorithm}

The $\pm J$ spin glass model is defined by the Hamiltonian
\be 
H \equiv -\sum_{<i,j>}\sigma_i J_{ij} \sigma_j\ ,
\label{E-HAMILTON}
\ee 
where the spins $\sigma_i$ take values $\pm 1$. The nearest neighbor
quenched couplings $J_{i j}$ take values $\pm 1$ with equal probability.
The spins live in a cubic lattice containing $V=L^3$ sites. We have used
helicoidal boundary conditions in two directions and periodic in the
third one. The reason is because we wanted to check the special purpose
computer developed for this physical model \cite{CLU}.

As usual, for every realization of the bonds or sample two independent copies 
of the system are studied. The main quantity to be measured is the
overlap between the two copies with the same disorder, 
which acts like an order parameter for
this model. The overlap between two spin configurations 
$\sigma$ and $\tau$ is given by
\be
  q(\sigma,\tau)\equiv\frac{1}{V}\sum_i 
     \sigma_i   \tau_i \ ,
  \protect\label{E-QSUM}
\ee
which is usually denoted $q$. Using powers of this quantity one 
can compute different observables. The second and fourth power
are used to build the Binder parameter
\be
  g \equiv \frac12 
  \left[
  3-\frac{\overline{\langle q^4\rangle}}{\overline{\langle 
  q^2\rangle}^2} \right] \ ,
  \label{E-BINDER}
\ee
where $\langle(\cdots)\rangle$ stands for the thermal average for
a given realization of the bonds, and $\overline{(\cdots)}$ means
the average over the disorder. Since this quantity is dimensionless,
it obeys the finite size scaling law (near the critical point)
\be
g = \tilde{g}\left( L^{1/\nu} \left( T-T_c\right) \right) \ ,
\label{E-BINLAW}
\ee
being independent of the volume at the critical temperature, $T=T_c$.
This property makes it appropriate to investigate the existence of 
any spin glass phase transition by studying the intersections of
the functions $g(T)$ for different lattice sizes. 

In addition, one can compute the spin glass susceptibility,
\be
  \chi \equiv  V{\overline{\langle q^2\rangle}}  \ .
  \label{E-SUSC}
\ee 

The algorithm we want to investigate is the demon algorithm
\cite{CREUTZ} proposed by Creutz. For this microcanonical algorithm,
the physical system is the standard lattice plus a demon, which acts
like an entity able to store energy. The update algorithm keeps
constant the sum of the energy of the lattice and the demon.

In order to carry out the MC simulation for a given total energy $H$,
one can start from a spin configuration with that energy and the demon
energy equal to zero. To generate new spin configurations, the spins
are updated as follows: first, a spin is selected and its sign is
proposed to be inverted. If the flip lowers the spin energy, the demon
takes that energy and the flip is accepted. On the other hand, if the
flip grows the spin energy, the change is only made if the demon has
that energy to give to the spin.

\begin{figure}[t]
\begin{centering}
\epsfig{figure=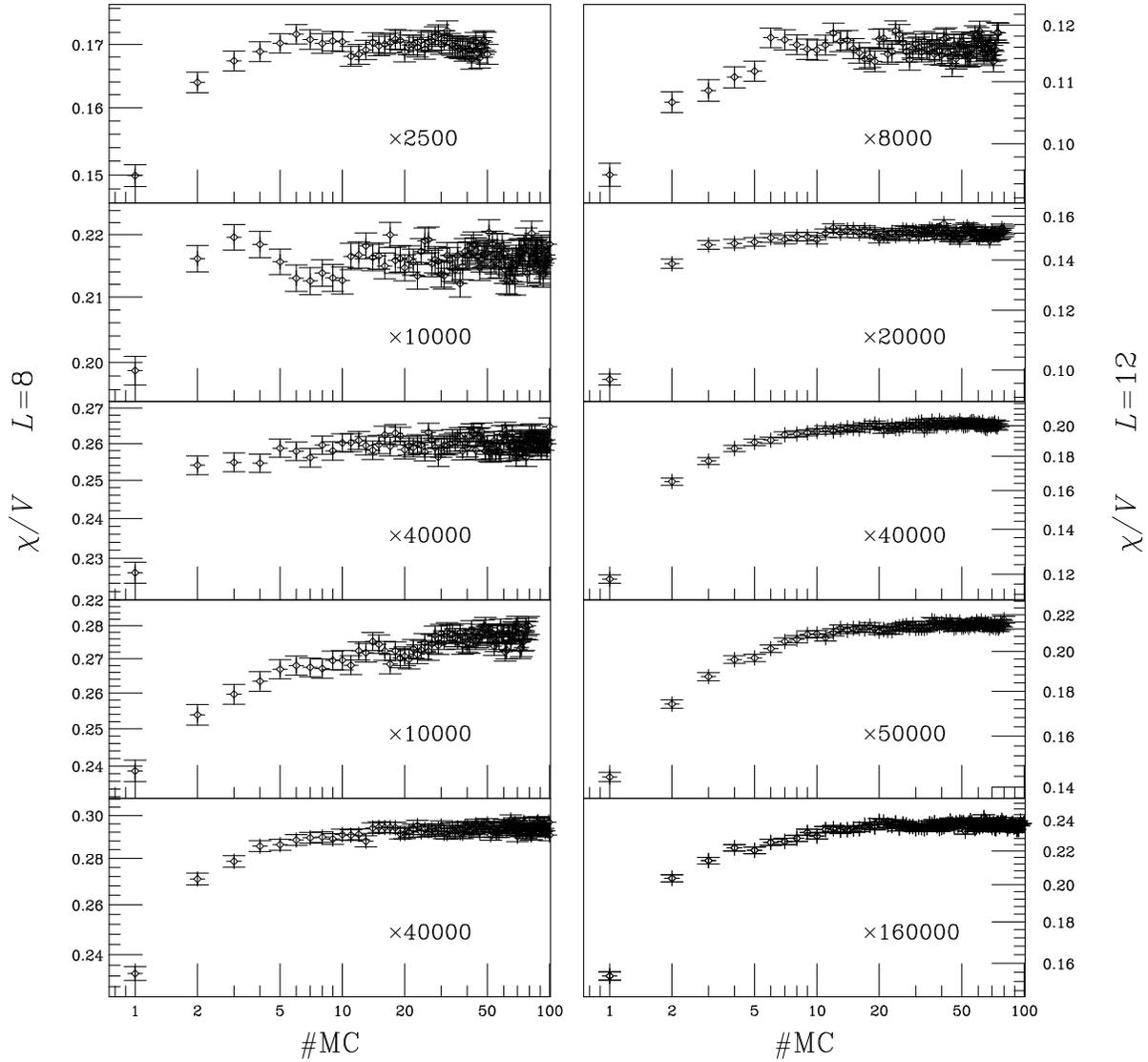,angle=90,width=160mm}
\vglue -0.5cm
\caption{Double-logarithmic plot of the MC evolution of $\chi/V$. From
top to bottom, $e= -1.650, -1.675, -1.700, -1.706$ and $-1.716$.
The factor is the scale of the $x$ axis.}
\label{F-logevol}
\end{centering}
\end{figure}

At this level, the value of the temperature $T$ is unknown and it can 
be obtained from the demon energy, whose probability distribution
$p(E_d)$ is given by the expression 
\be
p(E_d)\propto e^{-\beta E_d}.
\label{E-Ed1}
\ee

A fit to this function can provide the $\beta$ value, although 
a better estimate can be calculated if the mean energy of the 
demon on the sample, $\langle E_d \rangle$, is computed. Thus, 
the $\beta$ value is obtained as 
\be
\beta= \frac{1}{T} = \frac14\log \left( 1+\frac{4}{\langle E_d \rangle}\right).
\label{E-Ed2}
\ee

In a spin glass model, given the energy of the simulation, a different 
value of the temperature is obtained for every sample. The average
of them all gives the temperature of the simulation corresponding
to the fixed energy.

\begin{figure}[t]
\begin{centering}
\epsfig{figure=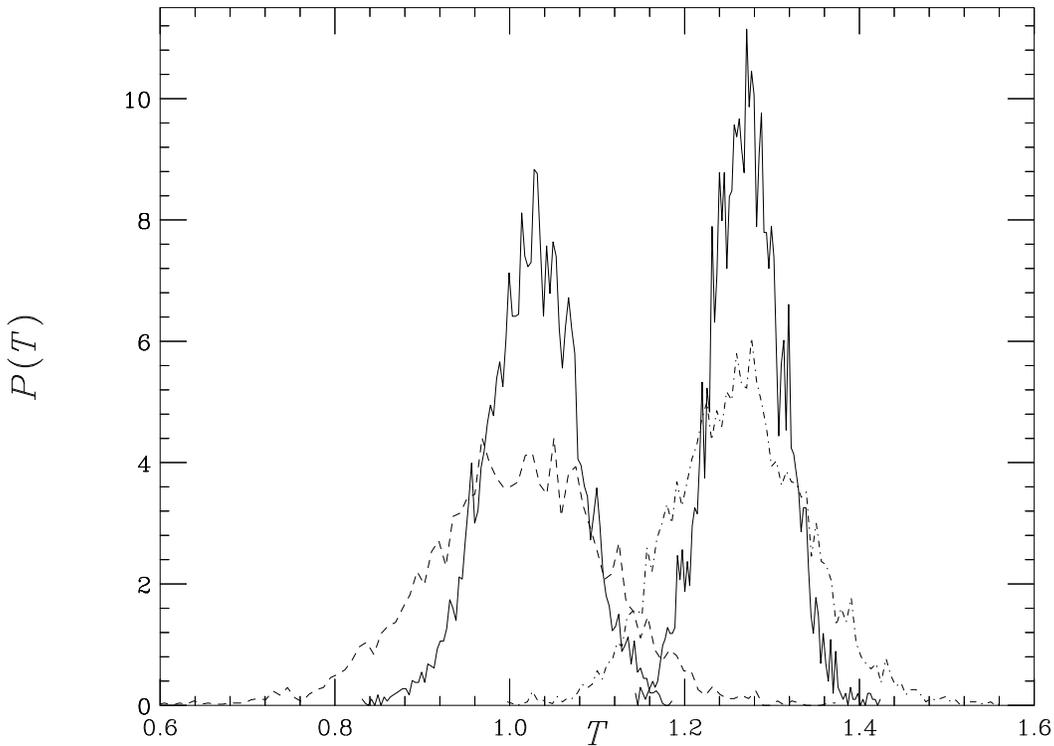,angle=90,width=140mm}
\vglue -0.5cm
\caption{Normalized distribution of temperatures for the highest
and lowest simulated energies in the $L=8$ (dashed lines) and
$L=12$ (solid lines) cases. The energies are $e=-1.716$ and $e=-1.650$}
\label{F-temperat}
\end{centering}
\end{figure}


\section{Numerical results}

We have simulated some different energies per spin $e$ in three lattice 
sizes ($L=8, 12$ and $16$). For every couple of parameters $(L,e)$ we generate 
a large set of samples. The initial demon energies and spin 
configurations are chosen to give a total energy $H=eV$. In order to 
generate them, we start with $E_d=0$ and all the spins equal to 1. We 
use the demon algorithm to change the spin configuration, but 
when $Ed>4$ we steal 4 units with a probability of 50\%. In this way, 
we approach smoothly towards the energy desired for the simulation. When 
the required energy is reached, the simulation starts. The spins
to be updated are sequentially chosen, having a completely updated spin
configuration after every $V$ updates.

\begin{table}[h]
\caption{{\it Parameters of the Microcanonicals runs }}
\label{T-STAT}
\centering
\small
\begin{tabular}{ccccccc}
\hline\hline
$L$ & $e$& Samples & \#MC sweeps &     $t_0$       &measures\\ \hline
8 & -1.650&3072&$1.25\times 10^5$& $2.5\times 10^4$& 50 \\ 
8 & -1.675&2048&$1.0\times 10^6$& $1.0\times 10^5$&100  \\ 
8 & -1.700&2048&$4.0\times 10^6$& $6.0\times 10^5$&300  \\ 
8 & -1.706&2304&$8.0\times 10^5$& $3.0\times 10^5$&300  \\ 
8 & -1.716&3328&$4.0\times 10^6$& $1.0\times 10^6$&250  \\ \hline
12& -1.650&1792&$6.0\times 10^5$& $2.0\times 10^5$&200  \\ 
12& -1.675&2048&$1.6\times 10^6$& $4.0\times 10^5$&400  \\ 
12& -1.700&3328&$3.0\times 10^6$& $1.0\times 10^6$&200  \\ 
12& -1.706&3584&$4.0\times 10^6$& $2.0\times 10^6$&200  \\ 
12& -1.716&4096&$1.6\times 10^7$& $6.4\times 10^6$&320  \\ \hline

16& -1.650&2560&$8.0\times 10^5$& $4.0\times 10^5$&200  \\ 
16& -1.700&2560&$4.0\times 10^6$& $1.5\times 10^6$&750  \\ \hline 
\hline
\end{tabular}
\normalsize
\end{table}

\begin{table}[h]
\caption{{\it Microcanonical Results}}
\label{T-RESULT}
\centering
\small
\begin{tabular}{cccccccc}
\hline
\hline
$L$ & $e$    & $T$ &  $\qcsq$&$\qcfo$  & $g$    \\ \hline
8 &-1.650&1.272(3)&0.169(1)&0.0453(4)&0.710(3)\\
8 &-1.675&1.182(3)&0.216(2)&0.0693(7)&0.759(4)\\
8 &-1.700&1.096(3)&0.260(2)&0.0966(9)&0.787(4)\\
8 &-1.706&1.059(2)&0.277(2)&0.108(1) &0.795(4)\\
8 &-1.716&1.025(2)&0.294(2)&0.121(1) &0.801(3)\\ \hline
12&-1.650&1.274(4)&0.116(1)&0.0222(3)&0.674(4)\\
12&-1.675&1.193(3)&0.152(1)&0.0358(4)&0.721(4)\\
12&-1.700&1.101(2)&0.201(1)&0.0587(5)&0.773(3)\\
12&-1.706&1.077(2)&0.214(1)&0.0657(5)&0.785(3)\\
12&-1.716&1.032(2)&0.237(1)&0.0793(6))&0.793(3)\\ \hline

16&-1.650&1.2765(30)&0.0841(7) &0.0122(2) &0.640(4)\\ 
16&-1.700&1.1033(2)  &0.1648(15)&0.04025(5)&0.7592(45)\\ \hline
\hline
\end{tabular}
\normalsize
\end{table}

Table \ref{T-STAT} shows the parameters of the different runs: lattice
size, energy per spin, number of samples and total number of Monte
Carlo sweeps are showed in the first columns. The next one shows the
thermalization time $t_0$ (we will discuss in detail below how we have
computed the thermalization time). The last column is the number of
measures of the overlap considered in every run to compute the
thermodynamical average.

The results obtained in these runs are shown in Table \ref{T-RESULT}.
We have computed the temperature according to Eq.~(\ref{E-Ed2}). The
second and fourth powers of the overlap have been also
calculated to obtain the Binder cumulant.  The work has been carried
out on the RTNN computer, which holds 32 PentiumPro processors, for a
total CPU time of approximately 20 days of
the whole machine. The errors in the estimates of the observables
have been calculated with the jack-knife method~\cite{jack-knife}.

\begin{table}[h]
\caption{{\it Parameters of the Metropolis runs }}
\label{T-CSTAT}
\centering
\small
\begin{tabular}{ccccccc}
\hline\hline
$L$ & $T$ & Samples & \#MC sweeps &     $t_0$     & measures \\ \hline
8 & 1.272&4000&$8.0\times 10^4$& $4.0\times 10^4$& 40   \\ 
8 & 1.182&4000&$2.0\times 10^5$& $7.0\times 10^4$& 130  \\ 
8 & 1.096&2390&$5.0\times 10^5$& $1.0\times 10^5$& 400  \\ 
8 & 1.059&4000&$2.0\times 10^5$& $1.4\times 10^5$&  60  \\ 
8 & 1.025&4000&$1.0\times 10^6$& $4.0\times 10^5$& 600  \\ \hline

12& 1.274&3700&$2.0\times 10^5$ & $1.2\times 10^5$& 80  \\ 
12& 1.193&2800&$2.0\times 10^5$ & $1.6\times 10^5$& 40  \\ 
12& 1.101&1328&$1.5\times 10^6$ & $7.0\times 10^5$&900  \\ 
12& 1.077&376 &$5.0\times 10^6$ & $1.5\times 10^6$&1000 \\ 
12& 1.032&1580 & $5.0\times10^6$ & $2.5\times 10^6$&2000   \\ 
\hline

16&1.2765&3520&$5.0\times 10^5$ & $4.2\times 10^5$& 80  \\ 
16&1.1033&2104&$4.0\times 10^6$ & $2.5\times 10^6$& 625 \\ \hline 
\hline
\end{tabular}
\normalsize
\end{table}

\begin{table}[h]
\caption{{\it Canonical Results}}
\label{T-CRESULT}
\centering
\small
\begin{tabular}{ccccccc}
\hline
\hline
$L$ &  $T$ &  $\qcsq$&$\qcfo$  & $g$ &$\Delta g$   \\ \hline
8 &1.272&0.170(1)&0.0481(5)&0.665(3)&     0.045(4) \\
8 &1.182&0.211(1)&0.0702(7)&0.714(3)&     0.045(5)   \\
8 &1.096&0.259(2)&0.099(1)&0.757(4)&      0.031(3)  \\
8 &1.059&0.275(2)&0.111(1) &0.769(4)&     0.010(5)  \\
8 &1.025&0.296(2)&0.1255(10) &0.7825(28)&  0.019(2)    \\ \hline

12&1.274&0.116(1)&0.0230(3)&0.645(5)& 0.032(6)\\
12&1.193&0.1497(17)&0.0362(6)&0.694(6)&0.027(7)\\
12&1.101&0.200(2)&0.059(1)&0.756(6)&0.017(6) \\
12&1.077&0.209(6)&0.065(3)&0.77(1)&0.015(10)\\
12& 1.032& 0.232(4)&0.078(1)&0.776(8) & 0.017(9)  \\ 
\hline
16&1.2765&0.085(1)&0.0128(2)&0.615(6)&0.025(7)\\
16&1.1033&0.161(2)& 0.0397(7) &0.733(7)& 0.026(8)\\ \hline
\hline
\end{tabular}
\normalsize
\end{table}

In order to be sure that the system is thermalized before
measuring, we check the symmetry on the probability distribution of the
overlap and also the MC evolution of the spin glass
susceptibility. For every run reported in this paper the mean value of
the overlap is zero (within the statistical error). Moreover, we have
checked the symmetry around zero of the probability distributions of
the overlap.

The MC evolution of the spin glass susceptibility is plotted in figure
\ref{F-logevol}. Every point in the plot has been computed by
averaging the values for the overlaps only in its corresponding MC
time. One expects to see the susceptibility rising with the Monte
Carlo time until a plateau is reached. The beginning of this plateau
defines the thermalization time $t_0$.  We use this criterion for the
thermalization.  As we said above, the temperature corresponding to
the simulation can be computed by using Eq.~\ref{E-Ed2}. For every
sample we obtain its own temperature.  Figure \ref{F-temperat} shows
the normalized probability distribution of temperatures obtained for
the highest and the lowest cases in both $L=8$ and $L=12$
lattices. Note the width of the histograms.

\section{Demon vs. Canonical: a comparison}

\begin{figure}[t]
\begin{centering}
\epsfig{figure=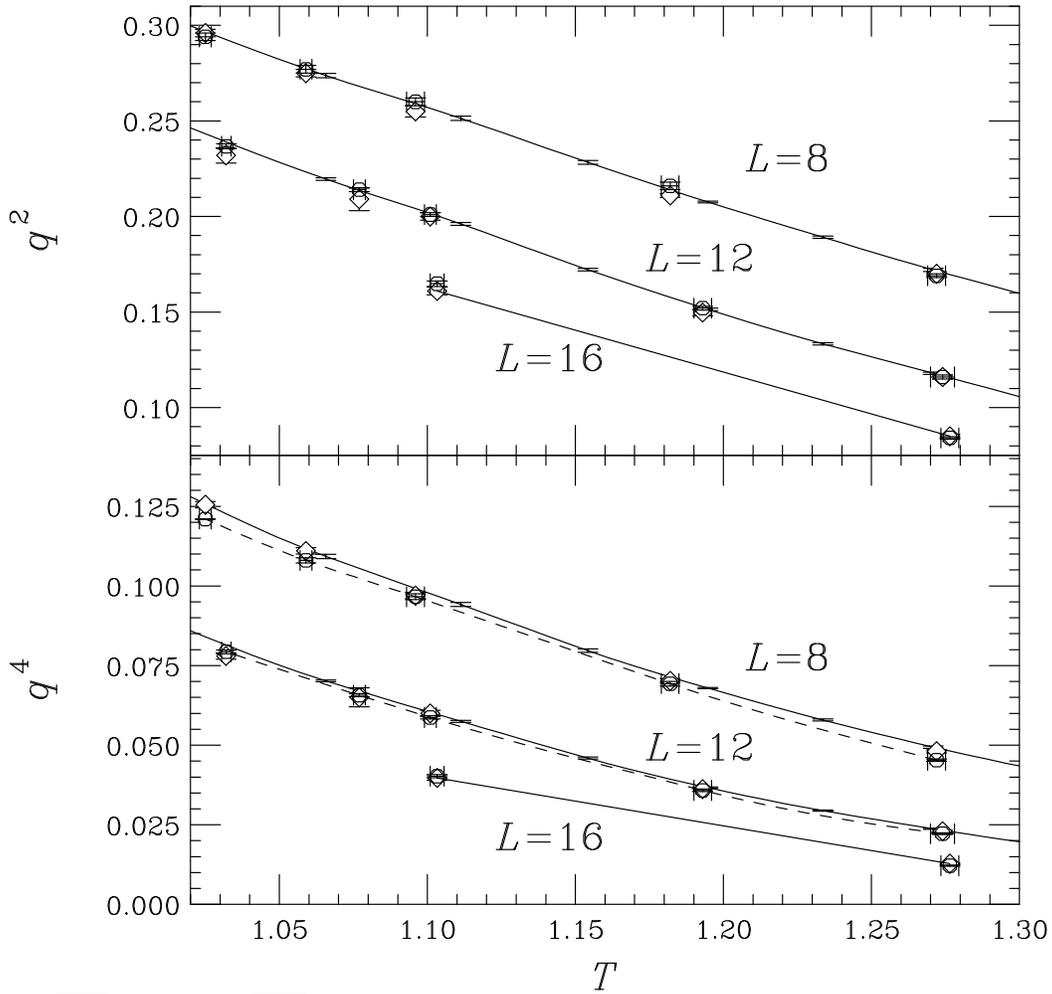,angle=0,width=140mm}
\vglue -0.5cm
\caption{$\qcsq$ (up) and $\qcfo$ (down) versus temperature for 
heat bath (dots), Metropolis (diamonds) and demon (circles)
algorithms. Lines are guides to the eyes.}
\label{F-powerofq}
\end{centering}
\end{figure}   

The results exposed in Table \ref{T-RESULT} can be compared with those
previously obtained \cite{YOUNG} in a canonical simulation running a
heat bath update algorithm. Moreover we have performed canonical
numerical simulations (using a Metropolis algorithm and periodic
boundary conditions) in order to run just at the temperature given by
the demon algorithm (and hence, compare at the same temperatures). We
report the parameters of these canonical simulations in Table
\ref{T-CSTAT} (the colums are: lattice size, temperature, number of
samples, number of Monte Carlo sweps, thermalization time ($t_0$) and
number of measures) and their results in Table \ref{T-CRESULT}. In
addition to this, in Table \ref{T-CRESULT} we have computed the
difference between the Binder cumulant computed in the demon
simulation and the canonical one.

Fig.~\ref{F-powerofq} shows the second and fourth moments of the
overlap. We take as reference the heat bath data from Kawashima and
Young~\cite{YOUNG} and our own Metropolis data. As a check of our
Metropolis simulation it is clear that our data match very well (one
standard deviation) between those from Kawashima and Young.

Now we can confront the demon data with the canonical.  While the
squared overlap seems to fit perfectly in the data obtained with heat
bath and Metropolis, the overlap to the fourth differs significantly,
being the microcanonical data lower than the canonical. The
discrepancy between the two ensembles decreases with the lattice size.

\begin{figure}[t]
\begin{centering}
\epsfig{figure=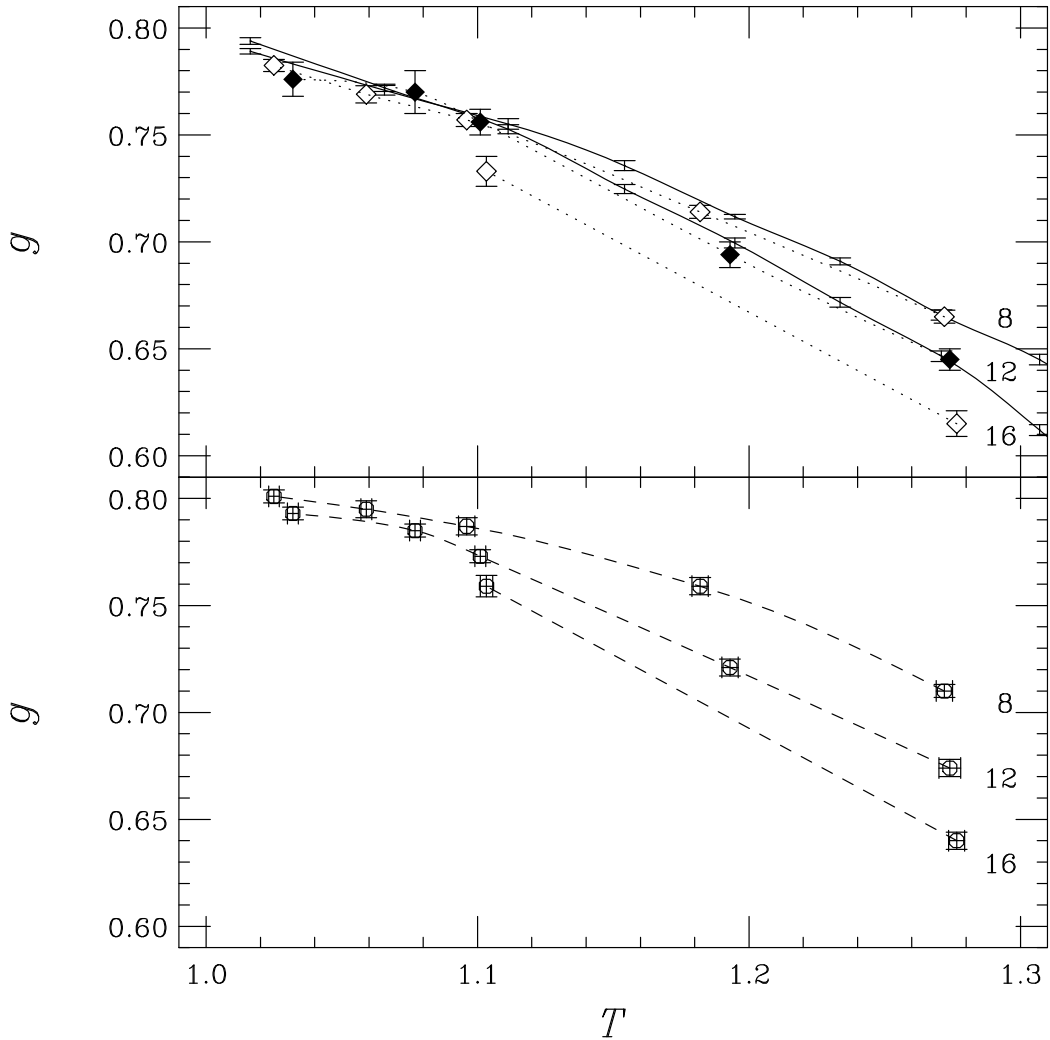,angle=0,width=140mm}
\vglue -0.5cm
\caption{Binder cumulant as a function of the temperature for
$L=8,12,16$. Heat bath (solid) and Metropolis (dotted) on the 
top,  demon (dashed) on the bottom.}
\label{F-binder}
\end{centering}
\end{figure}

These two quantities are used to calculate the Binder parameter, which
is plotted in Fig.~\ref{F-binder}.  The canonical simulation gives a
cut point between the two Binder parameters. In our microcanonical
simulation both Binder parameters approach at low temperature. Data
converge to be compatible in the error bar, but no cut point is
resolved using $L=8$ and $L=12$ data (we have simulated with the demon
algorithm $L=16$ data in the region $T \ge T_c=1.11$).

To be sure of the correctness of the algorithm we have carried out
some extra runs at $e=-1.650$ and $L=8$. One of them was using
periodic boundary conditions (to check the effect of periodic and
helicoidal boundary conditions on the observables using the demon
algorithm).  In addition, to check the ergodicity of the algorithm, we
repeated the simulation with the same realizations of the disorder but
starting from spin configurations obtained from a thermalized heat
bath simulation. In both cases, we obtained compatible results.

The canonical and microcanonical ensembles must agree in the
thermodynamical limit. The discrepancies between them have to decrease
when volume goes to infinite. To check it, we have simulated
$e=-1.650$ at $L=16$.  In this case, we obtained $g=0.640(4)$, nearer
to the Binder cumulant coming from the canonical simulation than the
$L=8$ and $L=12$ cases. We have seen that the discrepancy of the
Binder cumulant (sixth column of Table \ref{T-CRESULT}) for $e=-1.650$
goes to zero following a power law 
(with $\chi^2/{\rm d.o.f} \simeq 1$  where d.o.f stands for degrees
of freedom): $\Delta g (e=-1.65)
\propto L^{-0.86(36)}$. We can repeat this procedure for other
energies. For instance, if $e=-1.700$ then $\Delta g (e=-1.70) \propto
L^{-0.49(43)}$ with $\chi^2/{\rm d.o.f}=1.56$  
with a confidence level of 21\%.\footnote{The confidence
level is the  probability that $\chi^2$ were greater than the observed
value assuming that the statistical model used is correct (in our case
the power law behaviour). A very low value of this confidence level
(e.g. $<5\%$) would imply that our statistical model is incorrect.
See for instance~\cite{SOKAL}.} In any case, a more detailed study of
this issue is needed.

\begin{figure}[t]
\begin{centering}
\epsfig{figure=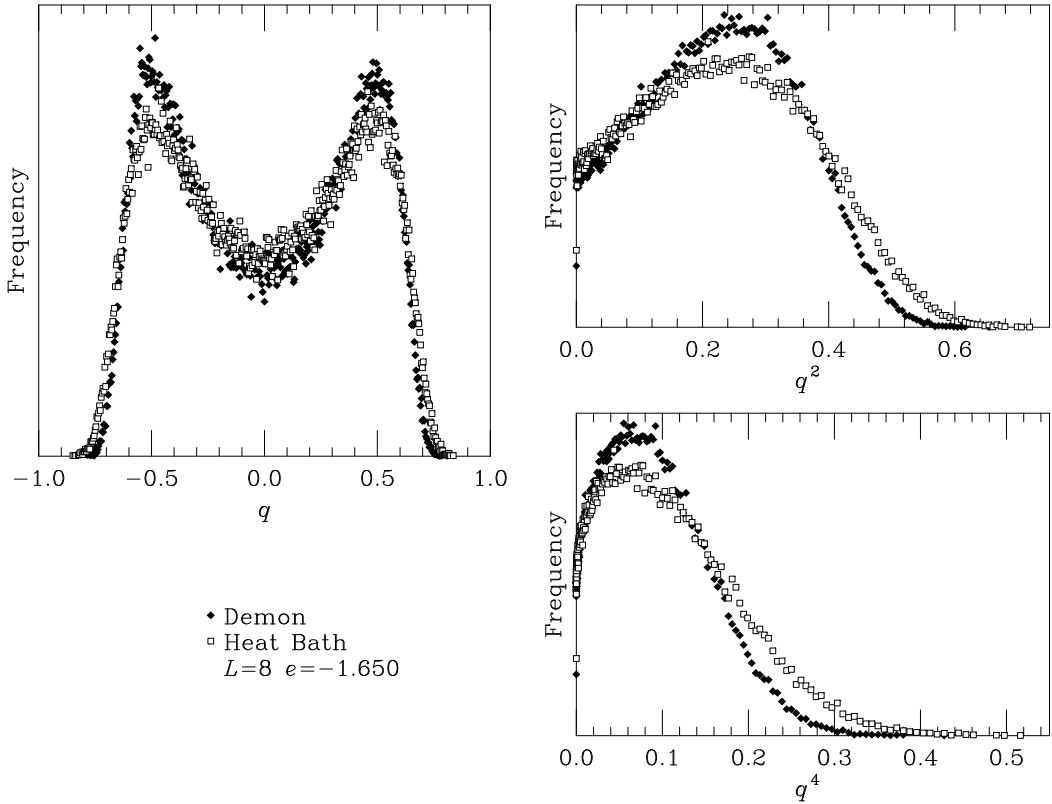,angle=90,width=140mm}
\vglue -0.5cm
\caption{Probability distributions at $T=1.272$, $L=8$.}
\label{F-dvshb}
\end{centering}
\end{figure}

We can study in more detail the previous issue by comparing the
probability distributions of the overlap, its second power, and its
fourth power obtained with the demon algorithm with $e=-1.650$.
Moreover, we have also measured the previous three probability
distributions carrying out a heat bath simulation at temperature
$T=1.272$ and $L=8$ with the same sets of bounds and number of
iterations of the case $e=-1.650$. We show these probability
distributions in Fig.~\ref{F-dvshb}.(a). The different shape of the
distribution is clarified in the plots of the powers of the overlap,
being the demon distribution more peaked than the canonical one.

\begin{figure}[t]
\begin{centering}
\epsfig{figure=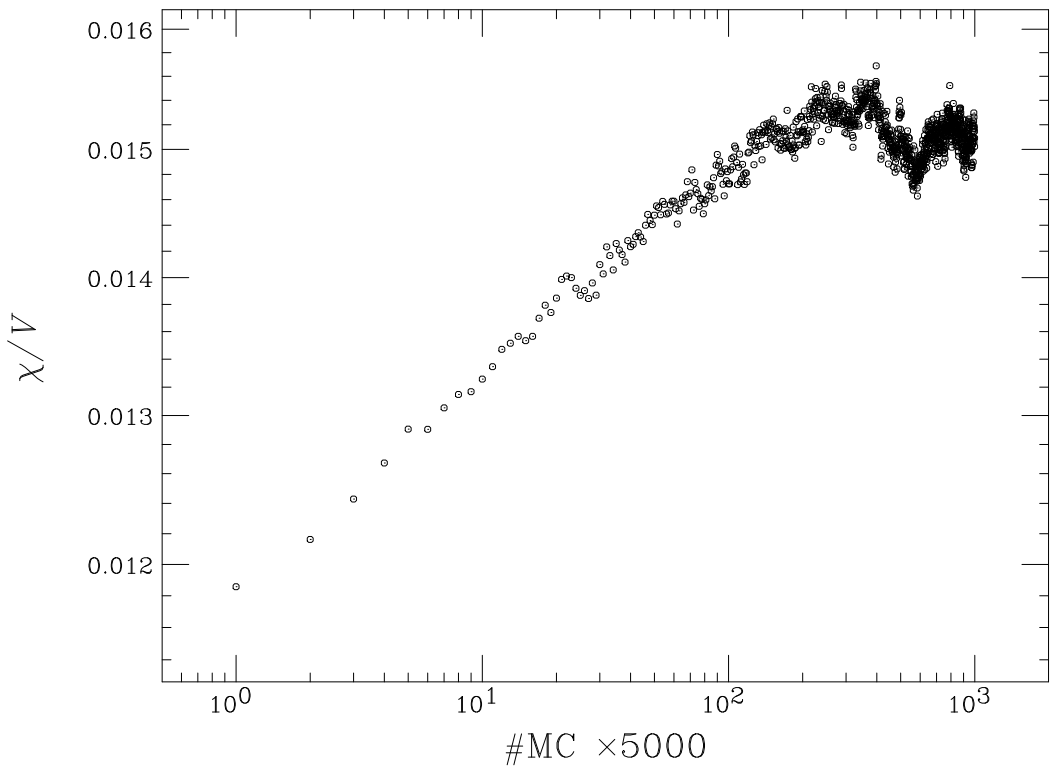,angle=90,width=140mm}
\vglue -0.5cm
\caption{Double-logarithmic plot of the MC evolution of $\chi/V$ for a 
canonical (Metropolis) simulation at $T=1.033$ in $L=12$.}
\label{F-logevolhb}
\end{centering}
\end{figure}

Other interesting issue is to compare the thermalization times needed
in the Metropolis simulation and in the demon algorithm. All these
thermalization times have been reported in the fifth column of Tables
\ref{T-STAT} and \ref{T-CSTAT}. We remark that these times have been
obtained by monitoring the growth of the non linear susceptibility as
a function of the Monte Carlo time. We take as $t_0$ (in the canonical
as well as in the microcanonical simulations) the Monte Carlo time in
which the numerical data achieve the equilibrium plateau (see
figures~\ref{F-logevol} and ~\ref{F-logevolhb} ). From the values of
$t_0$ for the lowest temperatures follows that demon algorithm
thermalizes slower (with a factor between 2 and 3) than
Metropolis. 

Obviously the cost of introducing a random number generator (i.e. to
run a canonical simulation instead of a microcanonical one) is less
than the factor two and three found in the autocorrelation times, and
so one of the conclusions of this paper is that (in an general purpose
computer) the efficiency of canonical algorithm are bigger than that of
the microcanonical one (bigger thermalization times).

It is clear that a dedicated machine with programmable logic (running
the microcanonical algorithm) with a speed 16 times bigger than a
supercomputer (running a Metropolis algorithm) clearly compensates
(for lattices of order $16$) the excess of thermalization time of the
demon algorithm (which is between 2 and 3). 
This is the situation if we compare in a {\em tower}
of APE100 supercomputer (which has a peak performance of 25
GigaFlops)~\cite{APE}, with a real performance of 5000 ps per spin.
and a machine with programmable logic (312 ps per spin). Of course, 
that special purpose computer running a Metropolis algorithm 
(at the same speed) would be even more efficient because the smaller
thermalization times. As a matter of fact, the special purpose 
computer referred in this work \cite{CLU} is able to run both 
algorithms at the same speed thanks to a fast random number generator 
implemented in hardware. The study of the efficiency of a combination
of the two algorithms is left to a later work.

We finally report our last check of the demon algorithm by checking
the Guerra relations which seem to be fulfilled within a 0.5\%
precision in a canonical simulation~\cite{MAPARURI}.

\begin{table}[t]
\small
\caption{{\it Guerra relation}.}
\label{T-GUERRA}
\centering
\begin{tabular}{cccccc}
\hline
\hline
$L$ & $e$    & $T$& lhs & rhs &lhs-rhs\\ \hline
8 &\-1.650&1.272(3)     &0.0331(6)&0.0351(5)&0.0020(8) \\ 
8 &\-1.675&1.182(3)     &0.0521(8)&0.0542(8)&0.0021(11) \\ 
8 &\-1.700&1.096(3)     &0.0748(9)&0.0774(9)&0.0026(13) \\ 
8 &\-1.706&1.059(2)     &0.0855(6)&0.0872(6)&0.0017(8) \\ 
8 &\-1.716&1.025(2)     &0.0960(9)&0.0978(9)&0.0018(13) \\ \hline
12&\-1.650&1.274(4)     &0.0155(2)&0.0163(2)&0.0008(3) \\ 
12&\-1.675&1.193(3)     &0.0265(3)&0.0272(3)&0.0007(4) \\ 
12&\-1.700&1.101(2)     &0.0458(5)&0.0465(5)&0.0007(10) \\ 
12&\-1.706&1.077(2)     &0.0518(7)&0.0525(7)&0.0007(10) \\ 
12&\-1.716&1.032(2)     &0.0632(8)&0.0638(8)&0.0006(11) \\ \hline
16&\-1.650&1.2765(30)   &0.0839(13)&0.0872(11)&0.0033(17) \\ 
16&\-1.700&1.1033(23)   &0.0323(4)&0.0312(4)&0.0011(6)  \\ \hline\hline
\end{tabular}
\normalsize
\end{table}

One of the Guerra's
relations~\cite{GUERRA} reads
\be
\overline{ {\langle q^2\rangle}^2 }  =
{1 \over 3}   \overline{\langle q^4\rangle} +
{2 \over 3}  {\overline{\langle q^2\rangle}}^2 \ .
\protect\label{E-GUERRA}
\ee

This relation has been shown exact for the Gaussian
model~\cite{GUERRA}. This relation can be rigourously demonstrated in
the infinite volume limit. However, one would expect finite
corrections in the Gaussian case.
Even though there is no proof for the $\pm J$
mode, the difference between the two sides of the equation has to
decrease with the volume. Table \ref{T-GUERRA} shows our results for
the left hand side (lhs) and the right hand side (rhs) of
Eq.~\ref{E-GUERRA}. The errors are calculated by a jack-knife
analysis.  

In the sixth column of Table \ref{T-GUERRA} we report the difference
between the lhs and the rhs of Eq.~\ref{E-GUERRA}. The maximum
deviation is 2.5 standard deviations ($L=8$ and $e=-1.650$). The rest
of the differences of the Table \ref{T-GUERRA} have fluctuations less
than two standard deviations. Hence, we can conclude that the Guerra
relation is satisfied in the demon algorithm.

\section{Conclusions}

We have studied a microcanonical algorithm running on the three
dimensional Ising spin glass in three dimensions.

We have obtained compatible (within the statistical errors) values
among the results of a canonical numerical simulation and the demon
algorithm for the second and fourth moments of the overlap whereas the
values of the Binder cumulant are different (but with the discrepancy
going to zero following a power law). We remark that microcanonical
and canonical algorithms should provide the same numerical results
only in the thermodynamic limit. Moreover the microcanonical algorithm
satisfies one of the Guerra relations. 

Finally we have shown that the thermalization times needed for the
demon algorithm are two or three times larger than those for the
Metropolis ones (for the larger simulated lattice $L=16$). 

We remark that we have checked numerically the ergodicity and the
efficiency of the algorithm.

From the point of view of the efficiency we have shown that the cost of
introducing random numbers is less than the excess of thermalization
which need the microcanonical simulation. Obviously, if we can design
a dedicated machine where only it is possible to implementate (via
hardware) a microcanonical algorithm, and if this dedicated machine
runs to a speed which is bigger than 5 times the speed of a canonical
code in a supercomputer the use of the microcanonical algorithm will
be welcome. Obviously this work shows that if we can implementate
random numbers with the cost of a factor two in time we should 
use the canonical algorithm instead of the microcanonical one.

We wish to thank J. M.~Carmona, L. A. Fern\'andez, D.~I\~niguez, G.
Parisi and A.~Taranc\'on for useful discussions. CLU is a DGA fellow. We 
also wish to thank P. Young for providing us the numerical
data of his reference~\cite{YOUNG}.

\end{document}